\documentclass{eptcs}
\providecommand{\event}{CLaC 2010} 
\usepackage{breakurl}             

\usepackage{amssymb}
\usepackage{amsmath}
\usepackage{bussproofs}
\usepackage{prooftree}
\usepackage{diagrams}

\usepackage{graphicx}

\newtheorem{lemma}{Lemma}
\newtheorem{definition}{Definition}
\newtheorem{proposition}{Proposition}
\newtheorem{corollary}{Corollary}

\newarrow{Map}{vee}---{>}
\newarrow{Rel}{}-{+}-{>}
\newarrow{Small}{}{}{}-{>}

\newcommand{\CL}{{\sf CL}}
\newcommand{\IL}{{\sf IL}}
\newcommand{\ML}{{\sf ML}}

\newcommand{\IT}[2]{I^{#1}_{#2}}
\newcommand{\Trans}[1]{{#1}^{Tr}}
\newcommand{\Tr}[1]{{#1}_{Tr}}
\newcommand{\G}[1]{{#1}^{G}}
\newcommand{\St}[1]{{#1}^{S}}
\newcommand{\GG}[1]{{#1}^{GG}}

\newcommand{\Ku}[1]{{#1}^{Ku}}
\newcommand{\Kup}[1]{{#1}^{\tilde{Ku}}}
\newcommand{\sKu}[1]{{#1}_{Ku}}

\newcommand{\Kr}[1]{{#1}^{Kr}}
\newcommand{\sKr}[1]{{#1}_{Kr}}
\newcommand{\Em}[1]{{#1}^{E}}
\newcommand{\E}[1]{{#1}_{E}}

\newcommand{\Ko}[1]{{#1}^{Ko}}

\newcommand{\T}[1]{{#1}^{T}}
\newcommand{\M}[1]{{#1}^{M}}
\newcommand{\Mp}[1]{{#1}^{M'}}

\newcommand{\CPS}{{\sf CPS}}

\newcommand{\pdefin}{:\equiv}
\newcommand{\eqleft}[1]{\begin{itemize} \item[] $#1$ \end{itemize}}

\newcommand{\proves}{\vdash}

\title{On Various Negative Translations}
\author{Gilda Ferreira \qquad \qquad Paulo Oliva 
\institute{Queen Mary University of London \\
School of Electronic Engineering and Computer Science \\ London, United Kingdom}
\email{\quad \qquad gilda@eecs.qmul.ac.uk  \quad\qquad paulo.oliva@eecs.qmul.ac.uk}}
\def\titlerunning{On Various Negative Translations}
\def\authorrunning{G. Ferreira \& P. Oliva}
\begin{document}
\maketitle

\begin{abstract}
Several proof translations of classical mathematics into intuitionistic mathematics have been proposed in the literature over the past century. These are normally referred to as \emph{negative translations} or \emph{double-negation translations}. Among those, the most commonly cited are translations due to Kolmogorov, G\"odel, Gentzen, Kuroda and Krivine (in chronological order). In this paper we propose a framework for explaining how these different translations are related to each other. More precisely, we define a notion of a (modular) simplification starting from Kolmogorov translation, which leads to a partial order between different negative translations. In this derived ordering, Kuroda and Krivine are minimal elements. Two new minimal translations are introduced, with G\"odel and Gentzen translations sitting in between Kolmogorov and one of these new translations.
\end{abstract}

\section{Introduction}\label{Intro}

With the discovery of paradoxes and inconsistencies in the early
formalisation of set theory, mathematicians started to worry about
the logical foundations of mathematics. Proofs by contradiction,
which concluded the existence of a mathematical object without actually
constructing it, were immediately thought by some to be the source
of the problem. Mathematicians were then segregated between those
who thought classical reasoning should be allowed as long as it was
finitistically justified (e.g. Hilbert) and those who thought proofs
in mathematics should avoid non-constructive arguments (e.g.
Brouwer). Constructivism and intuitionistic logic were born.

It was soon discovered, however, that the consistency of arithmetic
based on intuitionistic logic (Heyting arithmetic) is equivalent to
the consistency of arithmetic based on classical logic (Peano
arithmetic). Therefore, if one accepts that intuitionistic
arithmetic is consistent, then one must also accept that classical
arithmetic is consistent. That was achieved via a simple translation
of classical into intuitionistic logic which preserves the
statement $0 = 1$. So any proof of $0 = 1$ in Peano arithmetic (if ever one is found) can
be effectively translated into a proof of $0 = 1$ in Heyting
arithmetic.

The first such translation is due to Kolmogorov
\cite{Kolmogorov(25)} in 1925. He observed that placing a double
negation $\neg \neg$ in  front of every subformula turns a
classically valid formula into an intuitionistically valid one.
Formally, defining
\[
\begin{array}{cclccl}
\Ko{(A \wedge B)} & \; \pdefin \; & \neg \neg (\Ko{A} \wedge \Ko{B}) &
 \Ko{P} & \; \pdefin \; & \neg \neg P, \textup{~for $P$ atomic} \\
\Ko{(A \vee B)} & \pdefin & \neg \neg (\Ko{A} \vee \Ko{B}) &
 \Ko{(\forall x A)} & \pdefin & \neg \neg \forall x \Ko{A}\\
\Ko{(A \to B)} & \pdefin & \neg \neg (\Ko{A} \to \Ko{B}) \quad \quad &
 \Ko{(\exists x A)} & \pdefin & \neg \neg \exists x \Ko{A},
\end{array}
\]
one can show that $A$ is provable classically if and only if
$\Ko{A}$ is provable intuitionistically. Kolmogorov's translation,
however, was apparently not known to G\"odel and Gentzen who both
came up with similar translations
\cite{Gentzen(33),Gentzen(36),Goedel(33)} a few years later.
Gentzen's translation (nowadays known as G\"odel-Gentzen negative
translation \cite{Avigad(98),Ishihara(2000),Shirahata(2006)}) simply places a double negation in front of
atomic formulas, disjunctions, and existential quantifiers, i.e.
\[
\begin{array}{cclccl}
\GG{(A \wedge B)} & \; \pdefin \; & \GG{A} \wedge \GG{B} &
 \GG{P} & \; \pdefin \; & \neg \neg P, \textup{~for $P$ atomic} \\
\GG{(A \vee B)} & \pdefin & \neg \neg (\GG{A} \vee \GG{B}) \quad \quad &
 \GG{(\forall x A)} & \pdefin & \forall x \GG{A}\\
\GG{(A \to B)} & \pdefin & \GG{A} \to \GG{B}  &
 \GG{(\exists x A)} & \pdefin & \neg \neg \exists x \GG{A}.
\end{array}
\]
As with Kolmogorov's translation, we also have that $\CL \proves A$ if and only if $\IL \proves \GG{A}$, where $\CL$ and $\IL$ stand for classical and intuitionistic logic, respectively. G\"odel's suggested translation was in fact somewhere in between Kolmogorov's and Gentzen's, as it also placed a double negation in front of the clause for implication, i.e.
\eqleft{\GG{(A \to B)} \;\pdefin\; \neg (\GG{A} \wedge \neg \GG{B}) \; \Leftrightarrow_{\IL} \; \neg \neg (\GG{A} \to \GG{B}).}
In the 1950's, Kuroda revisited the issue of negative translations \cite{Kuroda(51)}, and proposed a different (somewhat simpler) translation:
\[
\begin{array}{cclccl}
\sKu{(A \wedge B)} & \; \pdefin \; & \sKu{A} \wedge \sKu{B} &
 \sKu{P} & \; \pdefin \; & P, \textup{~for $P$ atomic} \\
\sKu{(A \vee B)} & \pdefin & \sKu{A} \vee \sKu{B} \quad \quad &
 \sKu{(\forall x A)} & \pdefin & \forall x \neg \neg  \sKu{A}\\
\sKu{(A \to B)} & \pdefin & \sKu{A} \to \sKu{B}  &
 \sKu{(\exists x A)} & \pdefin & \exists x \sKu{A}.
\end{array}
\]
Let $\Ku{A} \pdefin \neg \neg \sKu{A}$. Similarly to Kolmogov,
G\"odel and Gentzen, Kuroda showed that $\CL \proves A$ if and only
if $\IL \proves \Ku{A}$. In particular, if $A$ does not contain
universal quantifiers then $\CL \proves A$ iff $\IL \proves \neg
\neg A$, since $\sKu{(\cdot)}$ is the identity mapping on formulas
not containing universal quantifiers. Finally, relatively recently,
following the work of Krivine \cite{Krivine(1990)}, yet another
different translation was developed\footnote{Throughout the paper
this translation is going to be called ``Krivine negative
translation'' as currently done in the literature (see
\cite{Streicher(2007),Kohlenbach(2008)}) even though it should be
better called Streicher-Reus translation. Although inspired by the
Krivine's work in \cite{Krivine(1990)} it is the syntactical
translation studied by Streicher and Reus \cite{Streicher(1998)} in
a version presented in \cite{Avigad(2006),Streicher(2007)} we are
using here.}, namely
\[
\begin{array}{cclccl}
\sKr{(A \wedge B)} & \pdefin & \sKr{A} \vee \sKr{B} &
 \sKr{P} & \pdefin & \neg P, \textup{~for $P$ atomic}\\
\sKr{(A \vee B)} & \pdefin & \sKr{A} \wedge \sKr{B} &
 \sKr{(\forall x A)} & \pdefin & \exists x \sKr{A} \\
\sKr{(A \to B)} & \pdefin & \neg \sKr{A} \wedge \sKr{B} \quad \quad &
 \sKr{(\exists x A)} & \pdefin & \neg \exists x \neg \sKr{A}.
\end{array}
\]
Letting $\Kr{A} \pdefin \neg \sKr{A}$, we also have that $\CL \proves A$ if and only if $\IL \proves \Kr{A}$.

It is also known that all these translations lead to
intuitionistically equivalent formulas, in the sense that $\Ko{A},
\GG{A}, \Ku{A}$ and $\Kr{A}$ are all provably intuitionistically
equivalent. As such, one could say that they are all essentially the
same. On the other hand, it is obvious that they are intrinsically
different.
The goal of the present paper is to explain the precise sense in
which G\"odel-Gentzen, Kuroda and Krivine translations are
systematic simplifications of Kolmogorov's original translation, and
show that, in a precise sense, the latter two are optimal (modular)
translations of classical logic into intuitionistic logic.
G\"{o}del-Gentzen translation is in between Kolmogorov's and a new
optimal variant we discuss in Section \ref{standard} below.

For more comprehensive surveys on the different negative translations,
with more historical background, see \cite{Kleene(52),Kohlenbach(2008),Luckhardt(73),Troelstra(73),Troelstra(96)}. \\[2mm]
{\bf Note}. Due to space restriction all proofs have been omitted. For all proofs see the full version of the paper at the authors webpages.

\subsection{Some useful results}

Our considerations on the different negative translations is based on the fact that formulas with various negations can be simplified to intuitionistically equivalent formulas with fewer negations. The cases when this is (or isn't) possible are outlined in the following lemma.

\begin{lemma}\label{EquivIL} The following equivalences are provable in $\IL$:
\[
\begin{array}{llll}
1. & \neg \neg (\neg \neg A \wedge \neg \neg B ) \leftrightarrow \neg \neg (A \wedge B) &
   9. & \neg \neg (\neg \neg A \wedge \neg \neg B)\leftrightarrow (\neg \neg A \wedge \neg \neg B) \\
2. & \neg \neg (\neg \neg A \vee \neg \neg B)\leftrightarrow \neg \neg (A \vee B)  &
   10. & \neg \neg (\neg \neg A \vee \neg \neg B)\leftrightarrow
   (\neg \neg \neg A \to \neg \neg B) \\
3. & \neg \neg (\neg \neg A \to \neg \neg B)\leftrightarrow \neg \neg (A \to B) \quad &
   11. & \neg \neg (\neg \neg A \to \neg \neg B)\leftrightarrow (\neg \neg A \to \neg \neg B) \\
4. & \neg \neg \exists x \neg \neg A \leftrightarrow \neg \neg \exists x A &
   12. & \neg \neg \forall x \neg \neg A \leftrightarrow \forall x \neg \neg A \\[2mm]
5. & \neg \neg (\neg A \wedge \neg B )\leftrightarrow \neg (A \vee B) &
   13. & \neg (\neg \neg A \wedge \neg \neg B )\leftrightarrow (\neg \neg A \to \neg B) \\
6. & \neg \neg (\neg A \vee \neg B)\leftrightarrow \neg (A \wedge B)  &
   14. & \neg (\neg \neg A \vee \neg \neg B)\leftrightarrow (\neg A \wedge \neg B) \\
7. & \neg \neg (\neg A \to \neg B)\leftrightarrow \neg ( \neg A \wedge B) &
   15. & \neg (\neg \neg A \to \neg \neg B)\leftrightarrow (\neg \neg A \wedge \neg B) \\
8. & \neg \neg \forall x \neg A \leftrightarrow \neg \exists x A &
   16. & \neg \exists x \neg \neg A \leftrightarrow \forall x \neg A.
\end{array}
\]
The following equivalences are provable in $\CL$ but not in $\IL$:
\[
\begin{array}{llll}
17. & \neg \neg \forall x \neg \neg A \leftrightarrow \neg \neg \forall x A &
   20. & \neg\neg \exists x \neg \neg A \leftrightarrow \exists x \neg \neg A \\
18. & \neg\neg \exists x \neg A \leftrightarrow \neg \forall x A &
   21. & \neg \forall x \neg \neg A \leftrightarrow \exists x \neg A \\
19. & \neg\neg (\neg\neg A \vee \neg \neg B) \leftrightarrow (\neg \neg A \vee \neg \neg B) \quad \quad &
   22. & \neg(\neg\neg A \wedge \neg \neg B) \leftrightarrow (\neg A \vee \neg B).
\end{array}
\]
\end{lemma}

\subsection{Logical framework}

In the language of classical logic $\CL$ and intuitionistic logic $\IL$, we consider
as primitive the constants $\bot$, $\top$, the connectives $\wedge$,
$\vee$, $\to$ and the quantifiers $\forall$ and $\exists$. We write $\neg A$ as an abbreviation for $A \to \bot$.
Note that $\CL$ can be formulated using a proper subset of the
symbols we consider as primitive. It would be sufficient, for
instance, to consider the fragment $\{\bot, \to, \vee, \exists\}$ or
$\{\bot, \to, \wedge, \forall\}$ (as adopted by Schwichtenberg in
\cite{Schwichtenberg(2006)}). Our choice of dealing directly with
the full set $\{\bot, \top, \to, \wedge, \vee, \forall, \exists\}$
in the classical framework has two main reasons: First, it
emphasises which symbols are treated in a similar or different
manner in classical and intuitionistic logic; second, in some
embeddings of $\CL$ into $\IL$ we are going to analyse, the
translations of certain formulas are syntactically different to the
derived translations we would obtain considering just a subset of
primitive symbols. In fact, usually when we choose to work with a
subset of the logical connectives in classical logic, we are
implicitly committing ourselves to one of the particular negative
translations.

\section{Modular Translations}

Let us first observe that all negative translations mentioned above
are in general not optimal -- in the sense of introducing the least
number of negations in order to turn a classically valid formula
into an intuitionistically valid one. For instance, Kuroda
translation of a purely universal formula $\forall x P(x)$ is $\neg
\neg \forall x \neg \neg P(x)$, whereas G\"odel-Gentzen would give
the optimal translation $\forall x \neg \neg P(x)$. On the other
hand, for purely existential formulas $\exists x P(x)$ we have that
Kuroda gives the optimal translation, whereas G\"odel-Gentzen
introduces unnecessary negations. The important property of all
these translations, however, is that they are \emph{modular}, i.e.
except for a single non-modular step applied to the whole formula,
the translation of a formula is based on the translation of its
immediate sub-formulas. The following definition makes this precise.

\begin{definition}[Modular negative translations] We say that a translation $\Trans{(\cdot)}$ from $\CL$ to $\IL$ is \emph{modular} if there are formula constructors $\IT{Tr}{\square}(\cdot, \cdot)$ for $\square\in \{\wedge, \vee ,\to \}$, $\IT{Tr}{Q}(\cdot,\cdot)$ for $Q\in \{\forall, \exists \}$, $\IT{Tr}{at}(\cdot)$ and $\IT{Tr}{\proves}(\cdot)$ called translation of connectives, quantifiers, atomic formulas and the provability sign, respectively, such that for each formula $A$ of $\CL$:
\eqleft{\Trans{A}\equiv \IT{Tr}{\proves}(\Tr{A})}
where $\Tr{(\cdot)}$ is defined inductively as:
\[
\begin{array}{cclccl}
\Tr{(A \wedge B)} & \pdefin & \IT{Tr}{\wedge}(\Tr{A}, \Tr{B}) &
 \Tr{P} & \pdefin & \IT{Tr}{at}(P), \textup{~for $P$ atomic}\\
\Tr{(A \vee B)} & \pdefin & \IT{Tr}{\vee}(\Tr{A}, \Tr{B}) &
 \Tr{(\forall x A)} & \pdefin & \IT{Tr}{\forall} (x,\Tr{A})\\
\Tr{(A \to B)} & \pdefin & \IT{Tr}{\to}(\Tr{A}, \Tr{B}) \quad \quad &
 \Tr{(\exists x A)} & \pdefin & \IT{Tr}{\exists}(x,\Tr{A}).
\end{array}
\]
A modular translation is called a \emph{negative translation} if (i)
$A \leftrightarrow_{\CL} \IT{Tr}{\proves}(\Tr{A})$ and (ii) $\IL
\proves \IT{Tr}{\proves}(\Tr{A})$ whenever $\CL \proves
A$.\footnote{A negative translation is usually assumed to satisfy a
third condition ($iii$) $\IT{Tr}{\proves}(\Tr{A})\leftrightarrow_{\IL}B$ for
some $B$ constructed from doubly negated atomic formulas by
means of $\forall, \wedge, \to, \bot$; ensuring that all negative
translations are equivalent (see \cite{Troelstra(73)}).}
\end{definition}
For instance, Krivine negative translation is a modular translation
with
\[
\begin{array}{cclccl}
\IT{Kr}{\wedge}(A, B) & \pdefin & A \vee B &
 \IT{Kr}{at}(P) & \pdefin & \neg P, \textup{~for $P$ atomic}\\
\IT{Kr}{\vee}(A, B) & \pdefin & A \wedge B &
 \IT{Kr}{\forall}(x,A) & \pdefin & \exists x A \\
\IT{Kr}{\to}(A, B) & \pdefin & \neg A \wedge B \quad \quad &
 \IT{Kr}{\exists}(x,A) & \pdefin & \neg \forall x \neg A
\end{array}
\]
and $\IT{Kr}{\proves}(A) \pdefin \neg A$. Similarly, one can easily
see how Kolmogorov, G\"odel-Gentzen, and Kuroda translations are
also modular translations.

\begin{definition}[Relating modular translations] We define a relation $\sim$ between modular translations as follows: Given translations $T_1$ and $T_2$ we define $T_1\sim T_2$ if the following equivalences are intuitionistically valid:
\[
\begin{array}{lcllcl}
\IT{T_1}{\square}(A,B) & \leftrightarrow_{\IL} & \IT{T_2}{\square}(A,B) \quad \quad &
   \IT{T_1}{at}(P) & \leftrightarrow_{\IL} & \IT{T_2}{at}(P) \\[2mm]
\IT{T_1}{Q}(x,A) & \leftrightarrow_{\IL} & \IT{T_2}{Q}(x,A) &
   \IT{T_1}{\proves}(A) & \leftrightarrow_{\IL} & \IT{T_2}{\proves}(A),
\end{array}
\]
for all formulas $A$, $B$, and atomic formulas $P$, $\square\in \{\wedge, \vee ,\to \}$ and $Q\in \{\forall, \exists
\}$.
\end{definition}

In other words, two modular translations are related via $\sim$ if the
corresponding translations of connectives, quantifiers, atoms and
provability are equivalent formulas in $\IL$.
%
It is immediate that $\sim$ is an \emph{equivalence relation}. In
what follows we say that two modular translations are the
\emph{same} if they are in the same equivalent class for the
relation $\sim$ (i.e. they are the same mod $\sim$). When two translations are not the same (in the previous sense), we
say they are \emph{different}. Two different translations $T_1$ and $T_2$ from $\CL$ to $\IL$ are said to be \emph{equivalent} if for each formula
$A$, the two translations of $A$, namely $A^{T_1}$ and $A^{T_2}$, are equivalent formulas in
$\IL$. For instance, changing the clause for $\exists x A$ in the G\"odel-Gentzen translation to $\GG{(\exists x A)} \pdefin \neg \forall x \neg \GG{A}$ does not change the interpretation, since intuitionistically we have that $\neg \forall x \neg A$ is equivalent to $\neg \neg \exists x A$.
So, these would be just two ways of writing the same translation.
On the other hand, Kuroda translation is different from
G\"odel-Gentzen's since, for instance, we do not normally have that
$\forall x A$ is  equivalent to $\forall x \neg \neg A$
intuitionistically.

\section{Simplifications}

Noticing that Kuroda and G\"{o}del-Gentzen negative translations
could be reached (in a modular way) from Kolmogorov translation via
equivalences in $\IL$, arose the idea of looking for a general
strategy covering the standard negative translations.

Thus, our goal is to show that the different negative translations
are obtained via a systematic simplification of Kolmogorov
translation. For that, we need the concept of ``simplification" we
define below. Intuitively, the idea of a simplification is to
transform formulas into intuitionistically equivalent formulas with
less negations \emph{preserving the modularity of the translation}.

\begin{definition}[Simplification from inside/outside] \label{simplification}
A \emph{simplification from inside} is a set of transformations (at most one for each connective and quantifier) of the following form:
\[
\begin{array}{ccl}
\neg \neg (N A \,\square\, N B) & \stackrel{r}{\Rightarrow} & N (N_1 A \;\square^{r} N_2 B) \\[2mm]
\neg\neg Q x N A & \stackrel{r}{\Rightarrow} & N (Q^{r} x N_1 A),
\end{array}
\]
where $\square, \square^{r} \in \{\wedge, \vee, \to\}$, and $Q,
Q^{r} \in \{\forall, \exists\}$, $N$ stands for a single or a double
negation (same choice in all the set of transformations), and
$N_1$ and $N_2$ are negations (possible none and not necessarily the
same in all transformations) such that
\begin{itemize}
\item[($i$)] both sides are equivalent formulas in $\IL$ and
\item[($ii$)] the number of negations on right side is strictly less than on left side.
\end{itemize}
A \emph{simplification from outside} is defined in a similar way replacing the shape of the transformation before by
\[
\begin{array}{ccl}
N(\neg \neg A \,\square\, \neg \neg B) & \stackrel{r}{\Rightarrow} & N_1 N A \, \square^{r} N_2 N B \\[2mm]
NQ x \neg \neg A & \stackrel{r}{\Rightarrow} & Q^{r} x  N_1 N A.
\end{array}
\]
\end{definition}

Intuitively, in the first case we are moving negations $N$
\emph{outwards} over the outer double negation $\neg \neg$, whereas
in the second case we are moving $N$ \emph{inwards} over the inner
$\neg \neg$. The moving of negations is done so that we reduce the
number of negations while keeping the modularity of the translation.

\begin{definition}[Maximal simplification] A simplification is \emph{maximal} if
\begin{itemize}
   \item[(i)] it is not properly included in any other simplification, i.e. including new transformations for other
connectives prevents the new set of being a simplification, and
   \item[(ii)] it is not possible to replace $\square^r$, $Q^r$, $N_1$ and $N_2$ so as to reduce the number of negations on the right side of any transformation.
\end{itemize}
\end{definition}

Intuitively, a simplification being maximal means that we can not get
ride of more negations.

\begin{proposition} \label{mrinside} Let $r_1$ and $r_2$ be the set of transformations:
\[
\begin{array}{lcllcl}
\neg \neg (\neg \neg A \wedge \neg \neg B) & \stackrel{r_1}{\Rightarrow} & \neg \neg (A \wedge B) &
   \neg \neg ( \neg A \wedge \neg B) & \stackrel{r_2}{\Rightarrow} & \neg ( A \vee B) \\
\neg \neg (\neg \neg A \vee \neg \neg B) & \stackrel{r_1}{\Rightarrow} & \neg \neg (A \vee B) &
   \neg \neg ( \neg A \vee \neg B) & \stackrel{r_2}{\Rightarrow} & \neg (A \wedge B) \\
\neg \neg (\neg \neg A \to \neg \neg B) & \stackrel{r_1}{\Rightarrow} & \neg \neg (A \to B) \quad \quad &
   \neg \neg (\neg A \to \neg B) & \stackrel{r_2}{\Rightarrow} & \neg (\neg A \wedge B) \\
\neg\neg \exists x \neg \neg A & \stackrel{r_1}{\Rightarrow} & \neg \neg \exists x A, &
   \neg \neg \forall x \neg A & \stackrel{r_2}{\Rightarrow} & \neg \exists x A,
\end{array}
\]
respectively. The sets $r_1$ and $r_2$ are maximal simplifications from inside.
\end{proposition}

\begin{proposition}\label{mroutside} Let $r_3$ and $r_4$ be the set of transformations:
\[
\begin{array}{lcllcl}
\neg \neg (\neg \neg A \wedge \neg \neg B) & \stackrel{r_3}{\Rightarrow} & \neg \neg A \wedge \neg \neg B &
   \neg (\neg \neg A \wedge \neg \neg B) & \stackrel{r_4}{\Rightarrow} & \neg \neg A \to \neg B \\
   \neg \neg (\neg \neg A \vee \neg \neg B) & \stackrel{r_3}{\Rightarrow} & \neg \neg \neg A \to \neg \neg B &
   \neg (\neg \neg A \vee \neg \neg B) & \stackrel{r_4}{\Rightarrow} & \neg A \wedge \neg B \\
   \neg \neg (\neg \neg A \to \neg \neg B) & \stackrel{r_3}{\Rightarrow} & \neg \neg A \to \neg \neg B \quad \quad &
   \neg (\neg \neg A \to \neg \neg B) & \stackrel{r_4}{\Rightarrow} & \neg \neg A \wedge \neg B\\
   \neg\neg \forall x \neg \neg A & \stackrel{r_3}{\Rightarrow} & \forall x \neg \neg A, &
   \neg \exists x \neg \neg A & \stackrel{r_4}{\Rightarrow} & \forall x \neg A,\\
\end{array}
\]
respectively. The sets $r_3$ and $r_4$ are maximal simplifications from outside.
\end{proposition}

\begin{proposition}\label{maxred} The simplifications $r_1$, $r_2$, $r_3$ and $r_4$ are the only maximal simplifications.
\end{proposition}

\section{Kolmogorov Simplified}

Definition \ref{simplification} identifies a class of
transformations which can be applied to Kolmogorov negative
translation without spoiling the modularity property of the
translation. We now present standard ways of simplifying Kolmogorov
translation via the maximal (or proper subsets of the maximal)
simplifications introduced above.

\begin{definition}[Simplification path] \emph{Applying a simplification to a formula $A$} consists in changing
the formula through successive steps, applying in each step a
transformation allowed by the simplification (i.e. transforming a
subformula having the shape of the left-hand side of the
transformation by the corresponding right-hand side), till no longer
be possible to simplify the expression via that simplification. We call
the path of formulas starting in $A$ we obtain this way a
\emph{simplification path}.
\end{definition}

Note that every step in a simplification path acts over a particular
connective or quantifier and all formulas in a simplification path are
equivalent formulas in $\IL$. The process of applying a simplification is
not unique and can lead to different formulas. Nevertheless, all simplification
paths are obviously finite since in each step the number
of negations is decreasing. From now on, we consider that all
simplification paths start with formulas in Kolmogorov form (i.e.
formulas of the form $\Ko{A}$).

\begin{definition}[Length of simplification path] The length of a simplification path $P$, denoted $s(P)$, is the number
of steps in $P$, or equivalently the number of nodes in $P$ minus
one, where by node we refer to each formula in $P$.
\end{definition}

Clearly, it is not true that two simplification paths with the same length lead to
the same formula, i.e. have the same final node. For instance, consider applying simplification $r_1$ to the formula below in two different ways:
\begin{center}
\setlength{\unitlength}{10mm}
\begin{picture}(6.0,3.5)
%
\put(1.3,3.0){$\neg \neg (\neg \neg (\neg \neg A \wedge \neg \neg B)\wedge \neg \neg \exists x \neg \neg A)$}
\put(-3,1.7){$\neg \neg (\neg \neg (\neg \neg A \wedge \neg \neg B)\wedge \neg \neg \exists x A)$}
\put(4,1.7){$\neg \neg (\neg \neg (A \wedge B)\wedge \neg \neg \exists x \neg \neg A)$}
\put(-2,0.2){$\neg \neg ( (\neg \neg A \wedge \neg \neg B)\wedge \exists x A)$}
\put(4,0.2){$\neg \neg ((A \wedge B)\wedge \exists x \neg \neg A)$}
\put(2,2.8){\line(-2,-1){1.0}}
\put(4,2.8){\line(2,-1){1.0}}
\put(1,1.5){\line(0,-1){0.8}}
\put(5,1.5){\line(0,-1){0.8}}
\end{picture}
\end{center}
Nevertheless, we prove that if a simplification is maximal or is a
subset of a maximal simplification then the length of the longest
paths is determined by the initial formula and, moreover, all the
paths with longest length lead to the same formula. In other words,
we have a kind of confluence property for longest paths. First some definitions and auxiliary results. \\[2mm]
{\bf Notation}. In order to simplify the formulation of Lemmas
\ref{Lemmar1} and \ref{Lemmar2r3r4} we use the following
abbreviations
\begin{itemize}
 \item \emph{Removing the double negations from inside} over $\square$ or $Q$, with $\square \in \{ \wedge,\vee, \to\}$ and $Q \in \{\forall,
\exists\}$, stands for replacing $\neg \neg (\neg \neg A \square
\neg \neg B)$ by $\neg \neg (A \square B)$, or $\neg \neg Q x \neg \neg A$ by $\neg \neg Q x A$.

 \item \emph{Removing the double negation from outside} over $\square\in \{\wedge, \to \}$ or $Q$ consists in replacing the formula $\neg \neg (\neg \neg A \square \neg \neg
B)$ by $\neg \neg A \square \neg \neg B$, or replacing $\neg \neg Q
x \neg \neg A$ by $Q x \neg \neg A$.

\item \emph{Removing the double negation from outside} over $\vee$ consists in replacing $\neg \neg (\neg \neg A \vee \neg \neg
B)$ by $\neg \neg \neg A \to \neg \neg B$.

 \item \emph{Removing single negations (from inside or outside)} over $\square\in \{\vee, \to \}$ in the formula $\neg \neg (\neg \neg A
\square \neg \neg B)$ consists in transforming the double negations
in single negations, replacing $\square$ by $\wedge$ and in the case
$\square\equiv\to$ adding a negation before $A$. \emph{Removing a
single negation (from inside or outside)} over a quantifier symbol $Q$ in
the formula $\neg \neg Q x \neg \neg A$ consists in replacing the
double negations by single negations and replacing $Q$ by its dual.

 \item \emph{Removing a single negation from inside (respectively outside)} over $\wedge$ in the formula $\neg \neg (\neg \neg A \wedge \neg \neg
B)$ consists in replacing this formula by $\neg(\neg A \vee \neg B)$
(or replacing this formula by $\neg (\neg \neg A \to \neg B)$
respectively).
\end{itemize}

We denote by $\#^{A}_\square$ and $\#^{A}_Q$ the number of symbols
$\square$ and $Q$ respectively, occurring in the formula A. For the
sake of counting symbols, the negation symbols $\neg$ introduced by
the translations are considered as primitive, and hence do not
change the value of $\#^{A}_\to$. For example $(\#^{A}_\to) =
(\#^{\Ko{A}}_\to)$.

\begin{lemma} \label{Lemmar1}
For the simplification $r_1$ and for any formula $\Ko{A}$ there is a
simplification path $P_{r_1}$ from $\Ko{A}$ such that
\eqleft{s(P_{r_1}) = (\#^{\Ko{A}}_\wedge) + (\#^{\Ko{A}}_\vee) +
(\#^{\Ko{A}}_\to) + (\#^{\Ko{A}}_\exists)}
and the formula in the last node can be obtained from $\Ko{A}$
locating in this formula all the occurrences of conjunctions,
disjunctions, implications and existential quantifications and
removing at once all the double negations from inside these
connectives and quantifiers.

Any simplification $r'_1$ obtained from $r_1$ by removing one or
more transformations admits a similar result discounting and
disregarding the logical symbols in the left-hand side of the
transformations removed.
\end{lemma}

The (omitted) proof above in fact provides an algorithm to construct
a simplification path for the simplification $r$ with $r\equiv r_1$
or $r\equiv r'_1$. The simplification path from $\Ko{A}$ constructed
this way is called \emph{standard path for $r$}.

\begin{lemma}\label{Lemmar2r3r4}
For the simplifications $r_2$, $r_3$, $r_4$ and for any formula $\Ko{A}$,
there are simplification paths $P_{r_2}$, $P_{r_3}$, $P_{r_4}$ such that
\begin{itemize}
 \item[] $s(P_{r_2}) = (\#^{\Ko{A}}_\wedge) + (\#^{\Ko{A}}_\vee) + (\#^{\Ko{A}}_\to) + (\#^{\Ko{A}}_\forall)$,
 \item[] $s(P_{r_3}) = (\#^{\Ko{A}}_\wedge) + (\#^{\Ko{A}}_\vee) + (\#^{\Ko{A}}_\to) + (\#^{\Ko{A}}_\forall)$ and
 \item[] $s(P_{r_4}) = (\#^{\Ko{A}}_\wedge) + (\#^{\Ko{A}}_\vee) + (\#^{\Ko{A}}_\to) + (\#^{\Ko{A}}_\exists)$.
\end{itemize}
Moreover, in $P_{r_2}$ the last node can be obtained from $\Ko{A}$
removing at once the single negations from inside all the
conjunctions, disjunctions, implications and universal
quantifications; the formula in the last node in $P_{r_3}$ can be
obtained from $\Ko{A}$ by removing at once the double negations from
outside the conjunctions, disjunctions, implications and universal
quantifications; and the formula in the last node of $P_{r_4}$ can
be obtained from $\Ko{A}$ by removing at once the single negations
from outside the conjunctions, disjunctions, implications and
existential quantifications.

The result can be adapted in the expected way to simplifications
obtained from $r_2$, $r_3$ or $r_4$ by removing one or more
transformations.
\end{lemma}

Again, the proof above provides algorithms to construct
simplification paths for the simplifications $r_2$, $r_3$, $r_4$ and
its subsets. The simplification paths from $\Ko{A}$ constructed via
these algorithms are called \emph{standard paths}.

\begin{lemma}
If the simplification is a subset of a maximal one, in each step of
a simplification path we act over a connective or a quantifier
already occurring in the initial formula, and we never act twice
over the same connective or quantifier.
\end{lemma}

Note that, in the previous lemma, the hypothesis of considering just
subsets of maximal simplifications is essential. In the example
below we present a (non maximal) simplification from inside that
contradicts the lemma. Consider the simplification:
\[
\begin{array}{ccl}
\neg \neg (\neg A \wedge \neg B) & \Rightarrow & \neg (A \vee \neg \neg B)\\[2mm]
\neg \neg ( \neg A \vee \neg B) & \Rightarrow & \neg (A \wedge B).
\end{array}
\]
From $\neg \neg (\neg \neg A \wedge \neg \neg (\neg \neg B \wedge
\neg \neg C))$ we can construct the following two paths:
\begin{center}
\setlength{\unitlength}{10mm}
\begin{picture}(6.0,5.0)
%
\put(1.3,4.5){$\neg \neg (\neg \neg A \wedge \neg \neg (\neg \neg B\wedge \neg \neg C))$}
\put(-2.4,3.2){$\neg ( \neg A \vee \neg \neg \neg (\neg \neg B\wedge \neg \neg C))$}
\put(4,3.2){$\neg \neg (\neg \neg A \wedge\neg(\neg B\vee \neg \neg\neg C))$}
\put(1.3,1.7){$\neg (\neg A \vee \neg \neg(\neg B\vee \neg \neg \neg C))$}
\put(1.7,0.4){$\neg (\neg A \vee \neg (B\wedge \neg \neg C))$}
\put(2,4.3){\line(-2,-1){1.0}}
\put(4,4.3){\line(2,-1){1.0}}
\put(1,2.8){\line(2,-1){1.0}}
\put(5,2.8){\line(-2,-1){1.0}}
\put(3.3,1.5){\line(0,-1){0.6}}
\end{picture}
\end{center}
The two corollaries below are now immediate:

\begin{corollary}
For each formula $\Ko{A}$ and each simplification that is a subset
of $r_1$, $r_2$, $r_3$ or $r_4$, any simplification path from
$\Ko{A}$ has length smaller or equal to the length of the
corresponding standard path.
\end{corollary}

\begin{corollary}
If the simplification is a subset of a maximal one, two
simplification paths with the longest length lead to the same
formula.
\end{corollary}

The result above justifies the next definition:

\begin{definition}
Let $r$ be a subset of a maximal simplification and $\Ko{A}$ a
formula in Kolmogorov form. We denote by $r(\Ko{A})$ the formula in
the last node of a simplification path with longest length.
\end{definition}

\section{Standard Translations}
\label{standard}

Simplifying the Kolmogorov negative translation via the maximal
simplifications $r_1$ and $r_2$ we obtain exactly Kuroda and Krivine
negative translations.

\begin{proposition} \label{prop-kuroda}
$r_1(\Ko{A})\equiv \Ku{A}$ and $r_2(\Ko{A})\equiv \Kr{A}$.
\end{proposition}

This study concerning maximal simplifications led us not only to the
two standard negative translations above but also to the discovery
of two new minimal modular embeddings from $\CL$ to $\IL$. Consider the
translations described below:
\[
\begin{array}{cclccl}
\G{(A \wedge B)} & \pdefin & \G{A} \wedge \G{B} \quad \quad &
 \G{P} & \pdefin & \neg \neg P, \textup{~for $P$ atomic}\\
\G{(A \vee B)} & \pdefin & \neg \G{A} \to \G{B} &
 \G{(\forall x A)} & \pdefin & \forall x \G{A}\\
\G{(A \to B)} & \pdefin & \G{A} \to \G{B} &
 \G{(\exists x A)} & \pdefin & \neg \neg \exists x \G{A}
\end{array}
\]
which is like the $\GG{(\cdot)}$-translation except for the $\vee$-clause where only one negation (rather than two) is introduced, and
\[
\begin{array}{cclccl}
\E{(A \wedge B)} & \pdefin & \neg \E{A} \to \E{B} \quad \quad &
 \E{P} & \pdefin & \neg P, \textup{~for $P$ atomic}\\
\E{(A \vee B)} & \pdefin & \E{A} \wedge \E{B} &
 \E{(\forall x A)} & \pdefin & \neg \forall x \neg \E{A}\\
\E{(A \to B)} & \pdefin & \neg \E{A} \wedge \E{B} &
 \E{(\exists x A)} & \pdefin & \forall x \E{A}
\end{array}
\]
with $\Em{A} \pdefin \neg \E{A}$, which is similar to Krivine except that negations are introduced in the $\{\wedge, \forall\}$-clauses whereas Krivine introduces negations on the $\exists$-clause.

Immediately as a corollary of the next proposition, we have that the
translations $\G{(\cdot)}$ and $\Em{(\cdot)}$ are embeddings from
$\CL$ to $\IL$, different but equivalent to the standard embeddings
considered previously.

\begin{proposition}
$r_3(\Ko{A})\equiv \G{A}$ and $r_4(\Ko{A})\equiv \Em{A}$.
\end{proposition}

Let $r'_3$ be the (non-maximal) simplification we obtain from $r_3$
by removing the transformation $\neg \neg (\neg \neg A \vee \neg
\neg B)  \Rightarrow \neg \neg \neg A \to \neg \neg
B$. We can easily prove that $r'_3(\Ko{A})\equiv \GG{A}$. Thus,
G\"{o}del-Gentzen negative translation is strictly in between Kolmogorov and
the $\G{(\cdot)}$-translation.

\section{Final remarks}

We conclude with a few remarks on two other negative translations, some related work and other avenues for further research.

\subsection{On non-modular negative translations}

Working with \emph{modular} translations brings various benefits. For instance, we can prove properties of the translation by a simple induction on the structure of the formulas, and when applying the translation to concrete proofs this can be done in a modular fashion. On the other hand, if we allow a translation to be non-modular, we can of course construct simpler embeddings, i.e. we can simplify Kolmogorov negative translation even more, getting ride of more negations.

For example, consider the simplification $r_3$ followed by one more
transformation $\neg\neg \exists x \neg \neg A \Rightarrow \neg
\forall x \neg A$ to be applied, whenever possible, at the end of
the simplification path. As such we could first simplify $\neg \neg
(\neg \neg A \wedge \neg \neg \exists x \neg \neg B)$ using $r_3$ to
the formula $\neg \neg A \wedge \neg \neg \exists x \neg \neg B$ and
then apply the final simplification to obtain $\neg \neg A \wedge
\neg \forall x \neg B$. Although non-modular, these kind of
procedures also give rise to translations of classical into
intuitionistic logic.

Avigad \cite{Avigad(00)} presented a more sophisticated non-modular translation that results from a fragment of $r_1$,
avoiding unnecessary negations. More precisely, Avigad's M-translation is defined as:
\[
\begin{array}{cclccl}
\M{(A\wedge B)} & \pdefin & \neg\M{(\sim A \vee \sim B)} \quad &
  \M{P} & \pdefin & P , \textup{~for $P$ atomic} \\[1mm]
\M{(A\vee B)} & \pdefin & \M{A} \vee \M{B} &
  \M{\bar{P}} & \pdefin & \neg P \\[1mm]
\M{(\forall x A)} & \pdefin & \neg \M{(\exists x \sim A)} &
  \M{(\exists x A)} & \pdefin & \exists x \M{A},
\end{array}
\]
where in classical logic we consider the negations of
atomic formulas $\bar{P}$ as primitive and the formula $\sim A$ is
obtained from $A$ replacing $\wedge$, $\forall$, $P$ respectively by
$\vee$, $\exists$ and $\bar{P}$ and conversely. Avigad showed that
\begin{itemize}
   \item[(1)] $\proves_{\IL} \neg \M{A} \leftrightarrow \neg \St{A}$
   \item[(2)] If $\proves_{\CL} A$ then $\proves _{\IL} \neg \M{(\sim A)}$,
\end{itemize}
where $\St{A}$ stands for any of the standard equivalent
translations mentioned before such as G\"{o}del-Gentzen, Kolmogorov,
Kuroda or Krivine negative translation.

\begin{lemma}\label{Avigad} $\neg \M{(\sim A)}\leftrightarrow_{\IL}\neg \neg \M{A}$
\end{lemma}

Although translation $\M{(\cdot)}$, as presented by Avigad, is not
modular, notice that it can be equivalently written in a modular way
as
\[
\begin{array}{cclccl}
\Mp{(A\wedge B)} & \pdefin & \neg\neg \Mp{A} \wedge \neg \neg \Mp{B} &
  \Mp{P} & \pdefin & P , \textup{~for $P$ atomic} \\[1mm]
\Mp{(A\vee B)} & \pdefin & \Mp{A} \vee \Mp{B} &
  \Mp{\bar{P}} & \pdefin & \neg P \\[1mm]
\Mp{(\forall x A)} & \pdefin & \forall x \neg \neg \Mp{A} &
  \Mp{(\exists x A)} & \pdefin & \exists x \Mp{A},
\end{array}
\]
since $\M{(\forall x A)}\pdefin \neg \M{(\exists x \sim A)}\pdefin
\neg \exists x (\M{(\sim A)})\leftrightarrow_{\IL}\forall x \neg
\M{(\sim A)}\stackrel{\textup{L}\ref{Avigad}}{\leftrightarrow}_{\IL} \forall x \neg
\neg \M{A}$ and
\eqleft{
\begin{array}{lcl}
\M{(A \wedge B)}
  & \pdefin & \neg \M{(\sim A \vee \sim B)} \pdefin \neg (\M{(\sim A)}\vee \M{(\sim B)}) \\[1mm]
  & \leftrightarrow_{\IL} & \neg \M{(\sim A)}\wedge \neg \M{(\sim B)} \stackrel{\textup{L}\ref{Avigad}}{\leftrightarrow}_{\IL} \neg \neg \M{A}\wedge \neg \neg \M{B}.
\end{array}
}
The translation $\Mp{(\cdot)}$ can be obtained from Kolmogorov
negative translation via a non-maximal simplification, more
precisely the simplification $r_1$ (corresponding to Kuroda
translation) without the transformation $\neg \neg (\neg \neg A
\wedge \neg \neg B) \stackrel{r_1}{\Rightarrow} \neg \neg (A \wedge
B)$.

Avigad's translation $\M{(\cdot)}$ is a \emph{non-modular}
simplification of $\Mp{(\cdot)}$ since for universal
quantifications, for conjunctions and for provability we replace
$\neg \neg \M{A}$ by $\neg\M{(\sim A)}$ which, although equivalent,
has possibly less negations, as we see in the (omitted) proof of
Lemma \ref{Avigad}. Moreover, as pointed by Avigad in
\cite{Avigad(00)}, we can simplify the translation $\M{(\cdot)}$
even further defining $\M{(A \wedge B)}$ as being $\M{A}\wedge
\M{B}$. The corresponding modular version in this case is exactly
Kuroda negative translation.

\subsection{On G\"odel-Gentzen negative translation}

Although nowadays it is common to name the translation
$\GG{(\cdot)}$, presented in Section \ref{Intro}, by
\emph{G\"{o}del-Gentzen negative translation}, a few remarks should
be made at this point. The translations due to G\"{o}del and Gentzen (\cite{Goedel(33)} and
\cite{Gentzen(36)}, respectively) where introduced in the context of
number theory translating an atomic formula $P$ into $P$ itself.
Later Kleene \cite{Kleene(52)} considered the translation of the pure
logical part, observing that double-negating atomic formulas was necessary, since one does not have stability $\neg \neg P \to P$ in general.

Rigorously, Gentzen's original formulation instead of double
negating disjunctions and existential quantifiers used the following
intuitionistic equivalent definitions $\GG{(A\vee B)}\pdefin \neg
(\neg \GG{A} \wedge \neg \GG{B})$ and $\GG{\exists x A}\pdefin \neg
\forall x \neg \GG{A}$, since, as such, one can then work in the $\{\exists,\vee\}$-free fragment of intuitionistic logic.

Moreover, as pointed in Section \ref{Intro} already, G\"{o}del's
original double-negation translation differs from Gentzen's negative
translation in the way implication is treated. We can easily see,
however, that G\"{o}del's negative translation can be obtained from
Kolmogorov negative translation via the non-maximal simplification
consisting in $r'_3$ without the transformation $\neg \neg (\neg
\neg A \to \neg \neg B) \Rightarrow \neg \neg A \to
\neg \neg B$, being, therefore, more expensive in term of negations
than Gentzen's negative translation. Another non-maximal
simplification, more precisely $r'_3$ without the transformation
$\neg \neg (\neg \neg A \wedge \neg \neg B)
\Rightarrow \neg \neg A \wedge \neg \neg B$, leads
to Aczel's $(\cdot)^{N}$ variant \cite{Aczel(2001)}.

Finally, we observe that sometimes in Kolmogorov or G\"{odel}-Gentzen negative
translations, $\bot$ is transformed differently from the other
atomic formulas, not into $\neg\neg \bot$ but into $\bot$ itself.
This change is easily adapted to our framework, considering in the
modular definition of a translation an extra operator
$\IT{Tr}{\bot}(\bot)$ and defining $\Tr{\bot}\pdefin
\IT{Tr}{\bot}(\bot)$. Note that the translations where
$\IT{Tr}{\bot}(\bot)\pdefin \bot$ are the same as the ones with
$\IT{Tr}{\bot}(\bot)\pdefin \neg \neg \bot$, since
$\bot\leftrightarrow\neg \neg \bot$ in $\IL$.

\subsection{On intuitionistic versus minimal logic}

More than translating $\CL$ into $\IL$, it is well known that some negative translations produce embeddings of $\CL$ into
minimal logic $\ML$ (i.e. intuitionistic logic without ex-falso-quodlibet). More precisely
\[
\begin{array}{ccl}
\CL\proves A & \textup{iff} & \ML\proves A^*,
\end{array}
\]
where $* \in \{Ko, GG\}$, for instance. But for Kuroda negative
translation we just have $\CL\proves A$ iff $\IL\proves \Ku{A}$ (see
\cite{Troelstra(96)}). In our framework, this appears as no surprise
since the direct implication in the transformation
\[
\begin{array}{ccl}
\neg \neg (\neg \neg A \to \neg \neg B) & \stackrel{r_1}{\Rightarrow} & \neg \neg (A \to B)\\
\end{array}
\]
is valid in $\IL$ but not in $\ML$. All the other equivalences in Lemma \ref{EquivIL} are provable in minimal logic. We observe, however, that a small change in Kuroda negative translation
produces an embedding in $\ML$. More precisely, if we change in
$r_1$ the clause for implication to
\[ \neg \neg (\neg \neg A \to \neg \neg B) \stackrel{\tilde{r}_1}{\Rightarrow} \neg \neg (A \to \neg \neg B) \]
we obtain a non-maximal simplification (in $\IL$) which corresponds
to a modular translation $\Kup{(\cdot)}$ between Kolmogorov and
Kuroda negative translations. Since $\neg \neg (\neg \neg A \to \neg
\neg B)\leftrightarrow_{\ML}\neg \neg (A \to \neg \neg B)$ the
simplification $\tilde{r}_1$ is maximal in $\ML$. Therefore, the
negative translation $\Kup{(\cdot)}$ that inserts $\neg \neg$ in
($i$) the beginning of the formula, ($ii$) after each universal
quantifier, and ($iii$) in front of the conclusion of each
implication is such that $\CL\proves A$ iff  $\ML\proves \Kup{A}$.

\subsection{Other related work}

{\bf Strong monads}. Part of the present study could have been developed in a more
general context. Let $\sf{T}$ be a (logical operator having the
properties of a) strong monad and consider the translation
$\T{(\cdot)}$ that inserts $\sf{T}$ in the beginning of each
subformula. Assuming that $\T{({\sf T} A)} \leftrightarrow {\sf T}
\T{A}$ what we obtain is a translation of $\ML + ({\sf T} A \to A)$
into $\ML$. We name such embedding \emph{Kolmogorov
$\sf{T}$-translation}. It can be seen that all the transformations
in simplifications $\tilde{r}_1$ and $r'_3$ remain valid
equivalences in $\ML$ when we replace $\neg\neg$ by any strong monad
$\sf{T}$. Thus, from Kolmogorov $\sf{T}$-translation we can obtain,
by means of the previous simplifications, the corresponding Kuroda
($\ML$ variant) and G\"{o}del-Gentzen $\sf{T}$-translations. As
particular cases we have
\begin{itemize}
   \item ${\sf T} A \pdefin \neg \neg A$ (recovering the standard double-negation translations),
   \item ${\sf T} B \pdefin (B \to A) \to A$ (corresponding to Friedman $A$-translations \cite{Friedman(78)}),
   \item ${\sf T} A \pdefin \neg A \to A$ or ${\sf T} A \pdefin (A \to R) \to A$ (Peirce translations \cite{Escardo(2010)}).
\end{itemize}
As references on these more general embeddings see \cite{Aczel(2001),Escardo(2010)}. \\[1.5mm]
{\bf Semantical approaches}. In this paper we did not discuss
semantical approaches to the negative translations. Some
considerations concerning conversions between Heyting and Boolean
algebras whose valuation of formulas is related via negative
translations can be found in \cite{Goedel(86),Rasiowa(63)} and a
more abstract treatment of negative translations in terms of
categorical logic can be found in
\cite{Hyland(02)}.\\[1.5mm]
{\bf $CPS$ transformations}. There is a close connection between
negative translations and \emph{continuation passing style} ($\CPS$)
transformations. In the literature
\cite{Fujita(95),deGroote(94),Streicher(1998)}, we can find various
$\CPS$-translations from $\lambda \mu$-calculus into
$\lambda$-calculus that correspond (at the type level) to the
standard negative translations.
Since the $\CPS$ technique captures evaluation ordering for the
source language (such as call-by-name, call-by-value, call-by-need)
it would be interesting to see if our simplifications linking the
standard negative translations can be expressed and are meaningful
at the calculus reduction strategy level. See also Chapters 9 and 10
in \cite{Murthy(90)}. \\[1.5mm]
{\bf Linear logic}. Although not addressed in this paper, the
refined framework of linear logic with its exponentials can be
useful in the study of the negative translations. It would be
interesting to analyse our simplifications through the
refined lens of Linear Logic. For related references see \cite{Girard(91),Danos(97),Laurent(03)}. \\[2mm]
{\bf Acknowledgements}. The first author was partially supported by the
EPSRC (grant EP/H011803/1), FCT (grant SFRH/BPD/34527/2006 and project PTDC/MAT/104716/2008) and CMAF. The second author gratefully acknowledges support of the Royal Society (grant 516002.K501/RH/kk). We would like to thank Jaime Gaspar for suggestions and comments in an earlier version of this paper.

\bibliographystyle{eptcs} 

\end{document}